\documentclass[]{eptcs}
\usepackage{breakurl}             

\usepackage{msc}
\usepackage{amsfonts}
\usepackage{mathtools}
\DeclarePairedDelimiter{\ceil}{\lceil}{\rceil}
\usepackage{amsmath}
\usepackage{amsthm} 
\usepackage{algorithm}
\usepackage{algpseudocode}
\usepackage[caption=false]{subfig}
\usepackage{graphicx}
\usepackage{caption}

\newtheorem{mydef}{Definition}
\newtheorem{myprop}{Proposition}

\title{Worst-case Throughput Analysis for Parametric Rate and Parametric Actor Execution Time Scenario-Aware Dataflow Graphs}
\author{Mladen Skelin
\institute{Norwegian University of Science and Technology\\
Trondheim, Norway}
\and
Marc Geilen
\institute{Eindhoven University of Technology\\
Eindhoven, The Netherlands}
\and
Francky Catthoor
\institute{IMEC vzw.\\
Leuven, Belgium}
\and
Sverre Hendseth
\institute{Norwegian University of Science and Technology\\
Trondheim, Norway}
}
\def\titlerunning{Worst-case Throughput Analysis for PSADF Graphs}
\def\authorrunning{M. Skelin et al.}
\begin{document}
\maketitle

\begin{abstract}
Scenario-aware dataflow (SADF) is a prominent tool for modeling and analysis of dynamic embedded dataflow applications. In SADF the application is represented as a finite collection of synchronous dataflow (SDF) graphs, each of which represents one possible application behaviour or scenario. A finite state machine (FSM) specifies the possible orders of scenario occurrences. The SADF model renders the tightest possible performance guarantees, but is limited by its finiteness. This means that from a practical point of view, it can only handle dynamic dataflow applications that are characterized by a reasonably sized set of possible behaviours or scenarios. In this paper we remove this limitation for a class of SADF graphs by means of SADF model parametrization in terms of graph port rates and actor execution times. First, we formally define the semantics of the model relevant for throughput analysis based on (max,+) linear system theory and (max,+) automata. Second, by generalizing some of the existing results, we give the algorithms for worst-case throughput analysis of parametric rate and parametric actor execution time acyclic SADF graphs with a fully connected, possibly infinite state transition system. Third, we demonstrate our approach on a few realistic applications from digital signal processing (DSP) domain mapped onto an embedded multi-processor architecture.
\end{abstract}

\section{Introduction}
Synchronous dataflow (SDF) \cite{2lee:all} was introduced as a restriction of Kahn process networks (KPN) \cite{2kahn} to allow compile-time scheduling. The term \textit{synchronous} means \textit{static} or \textit{regular}.
Synchronous dataflow graphs
(SDFGs) are directed graphs where nodes are called \textit{actors} and edges
are called \textit{channels}. The numbers of data samples produced or consumed are
known at compile time. We refer to these data samples as \textit{tokens} and to the
token production and consumption numbers as \textit{rates}. 
Although SDF is very fitted to model regular streaming applications, it is due to its static nature, very lacking in its ability to capture the dynamic behaviour of modern streaming applications.
Therefore, a notable number of SDF extensions has been proposed over the years. Cyclo-static dataflow (CSDF) \cite{2bil:all} allows token production and consumption to vary between actor firings as long as the variation forms a certain type of a periodic pattern, while models such as parametrized synchronous dataflow (PSDF) \cite{2bhatt:all}, variable-rate dataflow (VRDF) \cite{2wigg}, variable-rate phased dataflow (VPDF) \cite{2wigg} and schedulable parametric dataflow (SPDF) \cite{2frad:all} introduce parametric rates. Scenario-aware dataflow (SADF) \cite{2thel:all} encodes the dynamism of an application by identifying a finite number of different behaviours called \textit{modes} or \textit{scenarios}. Each of the modes is represented by a single synchronous dataflow graph. The modes or scenarios can occur in known or unknown sequences. A finite state machine (FSM) is used to encode occurrence patterns. SADF is equiped with a technique that yields the tightest possible performance guarantees \cite{2geil:all}. The power of this technique lies in its ability to consider transitions over all possible scenario sequences as given by the FSM. Considering only the worst-case scenario, i.e. the scenario with the lowest throughput, without considering scenario transitions could be too optimistic. On the other hand, merging all application SDFGs into one SDFG where an actor takes the worst-case execution time over all SDFGs in SADF would be too pessimistic. This is due to the fact that subsequent iterations belonging to different scenarios may overlap in time, i.e. execute in a pipelined fashion. However, SADF is limited by its finiteness. It can only handle a reasonably sized set of application scenarios. 

To illustrate this, let us define an abstract parallel application consisting of a nested \textit{for} loop with parametric affine loop bounds:
\begin{verbatim}
ProcessData.A(out g, out h);

for (i=0; i<=g; i++){
   for (j=0; j<=h; j++){
      // Perform two tasks in parallel
      #region ParallelTasks
      // Perform two tasks in parallel
      Parallel.Invoke(() =>
         {
            ProcessData.B(i,j);
         }, // close first parallel action
         () =>
         {
            ProcessData.C(i,j);	
         } // close second parallel action
      ); // close Parallel.Invoke
      #endregion

      ProcessData.D(i,j);
   }
}
\end{verbatim}
The example application consists of 4 subtasks: {\tt ProcessData.A}, {\tt ProcessData.B}, {\tt ProcessData.C}
and {\tt ProcessData.D} with known worst-case execution times. Data parallelism is elegantly specified using the {\tt Parallel.Invoke} construct. Inside the {\tt Parallel.Invoke} construct, an \textit{Action} delegate is passed for each item of work. The application is mapped onto a multi-processor platform. The task assignment employed is purely static. In order to add complexity, we assume that the application executes in a pipelined fashion, i.e. more instances of the application can be active at the same time. Such an assumption introduces resource dependencies over subsequent activations of the application. In other words, a subtask of the $(i+1)^{\mathrm{th}}$ activation of the application might have to wait for a certain subtask of the $i^{\mathrm{th}}$ activation to complete and release the corresponding processing element. As specified by the example code, $g$ and $h$ can take different values during each application execution, i.e. they are data-dependant and are the result of input data processing performed by the subtask  {\tt ProcessData.A}. Let us assume we know that $g$ can take the value from the interval $\left[0, \frac{n}{2}\right]$ and $h$ can take the value from the interval  $\left[0, \frac{m}{2}\right]$. In that case, from a pure timing perspective, this application will exhibit as many behaviours as there are integer points in the rational 2-polytope $ P_{n,m} $ given by the set of constraints $ \{0\le \frac{1}{2}n, \; 0\le \frac{1}{2}m\} $. For $ n=4500$ and $m=2001 $, to be able to use SADF to derive the tightest worst-case performance  bounds, even for such a simple application executing in a pipelined fashion on a multi-processor platform, we would have to generate $ 2,252,126 $ SDFGs \cite{2caluss:all}. The situation gets even worse when dealing with platforms that support dynamic voltage and frequency scaling (\textit{DVFS}), which is a commonly used technique that adapts both voltage and frequency of the system in respect to changing workloads \cite{2mack:all}. In this case also the execution times of the application subtasks would vary depending on the current DVFS setting of the processing element they are mapped to.

In our work, we will remove these limitations which hamper the use of SADF in important application domains. For this purpose, we will add parametrization to the basic SADF modeling approach both in terms of parametric rates and parametric actor execution times given over a parameter space, which is a totally non-trivial extension because the current core of the SADF framework relies strongly on the constant nature of the rates and actor execution times. We raise the problem of SADF parametrization in the scope of existing parametric dataflow models. PSDF \cite{2bhatt:all} and SPDF \cite{2frad:all} are two semantically very similar models that provide a high level of generalization. We prefer SPDF due to syntactical convenience. By incorporating SPDF semantics into the definition of our parametric rate and parametric actor execution time SADF (PSADF), we show that the SPDF model can at run-time be treated as a special case of a SADF. We then derive a technique for worst-case throughput analysis for PSADF. We demonstrate our approach on a few realistic applications from the digital signal processing (DSP) domain.
\section{Related Work}
Throughput analysis of SDFGs is studied by many authors. Reference \cite{2gha:all} gives a good overview of the existing methods. Due to the static nature of SDF, these methods cannot be applied to any form of parametric dataflow. \cite{3gha:all} presents three methods for throughput computation for an SDFG where actor execution times can be parameters. However, the technique does not consider parametric rates and can only handle the static case, i.e. the graph cannot change parameter values during its execution. \cite{2geil:all} introduces the (max,+) semantics for the SADF model relevant for worst-case performance analysis, but is, as previously mentioned, practically limited to a reasonably sized set of scenarios. The most closely related work to ours can be found in \cite{2dam:all}. It combines the approaches presented in \cite{2geil:all} and \cite{3gha:all} and yields a technique that finds throughput expressions for an SADFG where actors can have parameters as their execution times. However, the (max,+) semantics introduced in \cite{2dam:all} can consider only parametric actor execution times and not parametric rates. A straightforward extension of \cite{2dam:all} to cover the case of parametric rates is not possible because it is not clear how to symbolically execute the graph in the presence of parametric rates. In the scope of rate parametric dataflow models \cite{2bhatt:all}\cite{2frad:all}, little attention has been given to the aspect of time. Two examples of parametric models that explicitly deal with time are VRDF \cite{2wigg} and VPDF \cite{2wigg}. These address the problem of buffer capacity computation under a throughput constraint, but both have a structural constraint that each production of $p$ tokens must be matched by exactly one consumption of $p$ tokens. That drastically limits the scope of applications it can consider.

So, the current approaches in throughput analysis for dataflow MoCs either cannot consider parametric rates \cite{2gha:all}\cite{3gha:all}\cite{2geil:all}\cite{2dam:all}, or impose too hard structural constraints that severely limit the expressivity of the model \cite{2wigg}. In our work we will remove these limitations by embedding the SPDF model \cite{2frad:all} which provides a high level of generalization into the SADF model \cite{2thel:all}\cite{2geil:all}.
\section{Preliminaries}
\subsection{Synchronous Dataflow Graphs}
%
%
%
%
%
%
%
SDFG is a directed graph $(\mathcal{A}, \mathcal{E})$ where nodes represent \textit{actors} which in turn represent functions or tasks, while edges represent their dependencies. We also refer to edges as \textit{channels}. Execution of an actor is denoted as firing and it is assigned with a time duration. In SDF, the number of tokens consumed and produced by an actor is constant for each firing. We refer to these numbers as \textit{rates}. Actors communicate using tokens sent over channels from one actor to another.
Fig. \ref{fig:example_sdf} shows an example of an SDFG with 5 actors ($\mathcal{A} = \{A,B,C,D,E\}$) and 9 channels ($\mathcal{E} = \{(A,B), (B,C), (C,C), \ldots\}$). Some channels might contain initial tokens, depicted with solid dots. The example graph contains 5 initial tokens that are labeled $t_1,\ldots,t_5$. Each actor is assigned with a firing time duration, denoted in the actor node, below the actor name, e.g. actor $A$ has a firing duration of $29$ time-units. Each port is assigned with a rate. When the value is omitted, it means that the value equals to $1$. 
As rates in SDF are constant for each firing, it is possible to construct a finite schedule (if it exists) that can be periodically repeated \cite{2lee:all}. Such a schedule assures liveness and boundedness \cite{2lee:all}. We call such minimal sequence of firings an iteration of the SDFG. This is a sequence of firings that has no net effect on the token distribution in the graph. The numbers of firings of each actor within an iteration constitute the \textit{repetition vector} of an SDFG. We only consider dataflow graphs that are bounded and live. Throughput is considered in terms of the number of iterations per time-unit, i.e. the number of iterations executed in one period normalized by the repetition vector divided by the duration of the period \cite{2gha:all}. It is natural to do so, because an iteration represents a coherent set of calculations, e.g. decoding of a video frame. For more details we refer to \cite{2lee:all}\cite{2gha:all}.
\subsection{(max,+) Algebra for SDFGs}
\begin{figure}[t]%
\centering
	\subfloat[][An example SDFG.]{
	\label{fig:example_sdf}
	\includegraphics[width=2.2in]{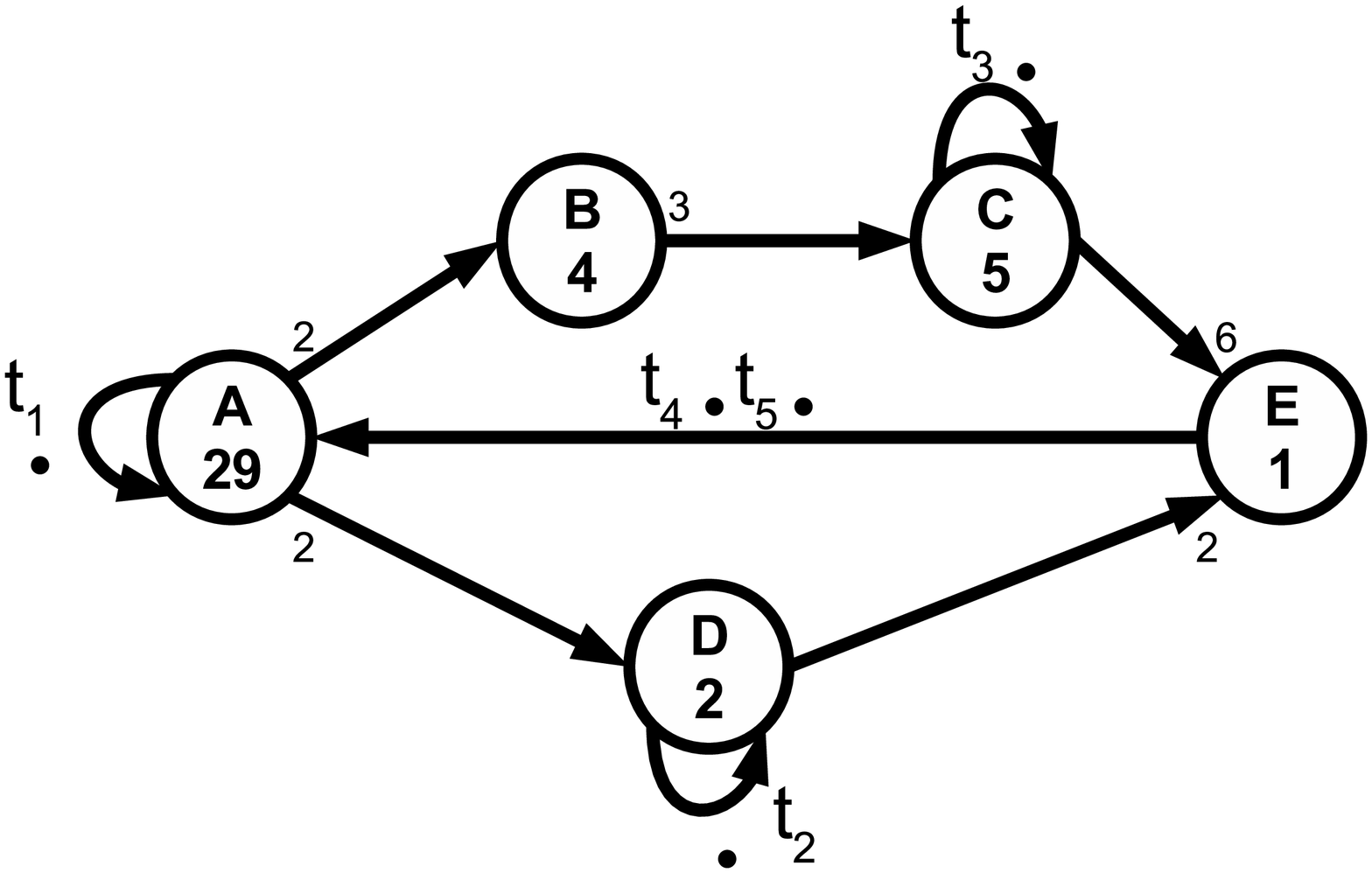}}%
	~~~~~~~~~~~~~~~
	\subfloat[][MPAG of the example SDFG.]{
	\label{fig:example_sdf_mpag}
	\includegraphics[width=1.3in]{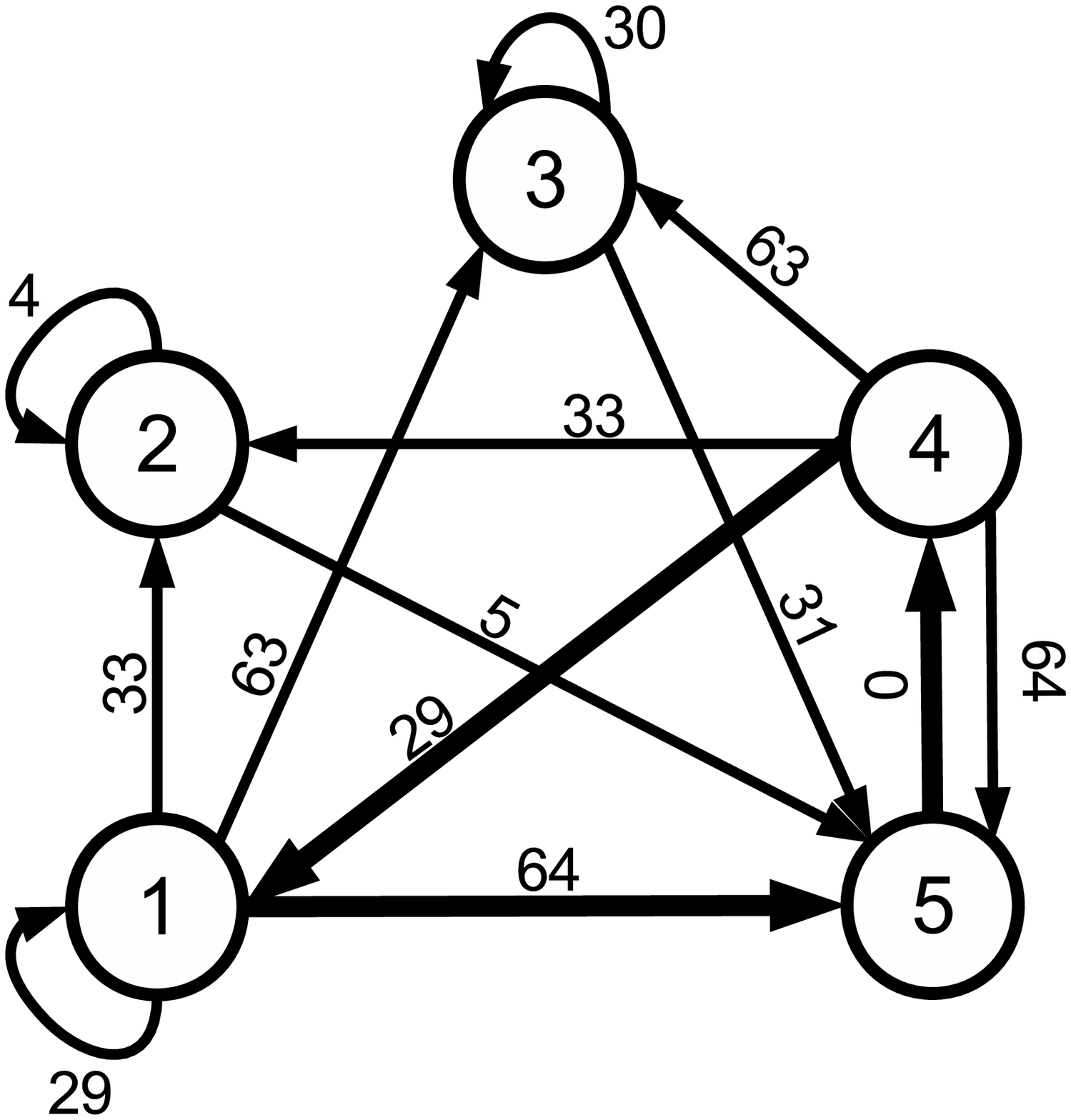}}
	\caption{Synchronous dataflow}%
	\label{fig:sdf}%
\end{figure}

Let $a \oplus b = \mathit{max}(a, b)$, $a \otimes b = a + b$ for $a, b \in \overline{\mathbb{R}} = \mathbb{R} \cup \{-\infty\}$. By max-algebra we
understand the analogue of linear algebra developed for the pair of operations $(\oplus,\otimes)$ extended
to matrices and vectors \cite{2bacc:all}.
Let $\vec{\gamma}$ denote the vector of production times of tokens that exist in their different channels in between iterations, i.e. it has an entry for each initial token in the graph. Then  $\vec{\gamma}_k$ denotes the vector of production times of initial tokens after $k$ iterations of the graph. These vectors then can be found using (max,+) algebra \cite{2bacc:all}. The evolution of the graph is then given by the following equation: $\vec{\gamma}_{k+1}=\mathbf{G} \vec{\gamma}_k$, where $\mathbf{G}=\{g_{ij}\}$ is a (max,+) characteristic matrix of the graph. Entry $g_{ij}$ specifies the minimal elapsed time from the production time of the $j^{\mathrm{th}}$ token in the previous iteration to the production time of the $i^{\mathrm{th}}$ token in the current iteration. When the $i^{\mathrm{th}}$ token is not dependent on the $j^{\mathrm{th}}$ token, then $g_{ij}=-\infty$. The specification of the algorithm for obtaining $\mathbf{G}$ can be found in \cite{2geil}. The (max,+) characteristic matrix for the example SDFG in Fig. \ref{fig:example_sdf} takes the form:
\[ \mathbf{G} = \begin{bmatrix}
29 & -\infty & -\infty & 29 & -\infty \\
33 & 4 & -\infty & 33 & -\infty\\
63 & -\infty & 30 & 63 & -\infty\\
-\infty & -\infty & -\infty & -\infty & 0\\
64 & 5 & 31 & 64 & -\infty\\
\end{bmatrix}.\]
For example, $\vec{\gamma}_1$ can be calculated as below:
\[ \vec{\gamma_1} = \begin{bmatrix}
29 & -\infty & -\infty & 29 & -\infty \\
33 & 4 & -\infty & 33 & -\infty\\
63 & -\infty & 30 & 63 & -\infty\\
-\infty & -\infty & -\infty & -\infty & 0\\
64 & 5 & 31 & 64 & -\infty\\
\end{bmatrix}
\begin{bmatrix}
0\\
0\\
0\\
0\\
0\\
\end{bmatrix}=
\begin{bmatrix}
\mathit{max}(29+0,29+0)\\
\mathit{max}(33+0, 4+0, 33+0)\\
\mathit{max}(63+0,30+0,63+0)\\
\mathit{max}(0+0)\\
\mathit{max}(64+0,5+0,31+0,64+0)\\
\end{bmatrix}=
\begin{bmatrix}
29\\
33\\
63\\
0\\
64\\
\end{bmatrix}
.\]

Paper \cite{2geil:all} explains how to obtain the throughput of an SDFG from the matrix $\mathbf{G}$. Briefly, matrix $\mathbf{G}$ defines a corresponding (max,+) automaton graph (MPAG) \cite{2gaub}. MPAG has as many nodes as there are initial tokens in the graph. An edge with the weight $g_{ij}$ is created from the $j^{\mathrm{th}}$ node to the $i^{\mathrm{th}}$ if $g_{ij} \ne -\infty$. The maximum cycle mean (MCM) $\lambda$ of the MPAG identifies the critical cycle of the SDFG. The critical cycle limits the throughput of the SDFG which takes the value $1 / \lambda$. MPAG of the example SDFG graph is displayed in Fig. \ref{fig:example_sdf_mpag}. The cycle with  weights $g_{14}-g_{51}-g_{45}$ (denoted with bold arrows) determines the throughput which takes the value of $1/31$ iterations per time-unit.
\subsection{Scenario-Aware Dataflow Graphs (SADFG)}
\begin{figure}[t]%
\centering
	\subfloat[][An example SADFG.]{
	\label{fig:example_ssdf}
	\includegraphics[width=1.9in]{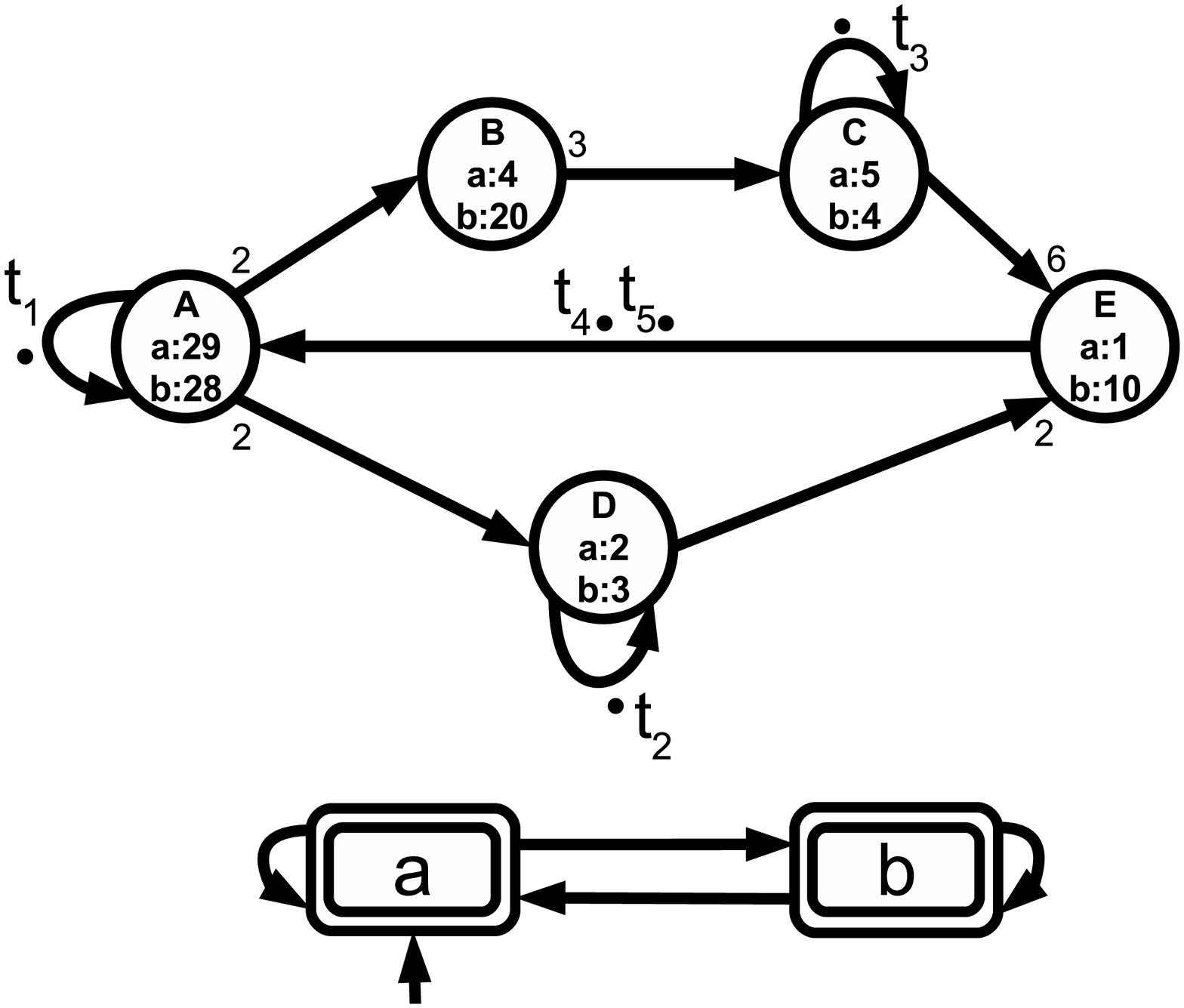}}%
	~~~~~~~~~~~~~~~
	\subfloat[][MPAG of the examle SADFG.]{
	\label{fig:example_sadf_mpag}
	\includegraphics[width=1.5in]{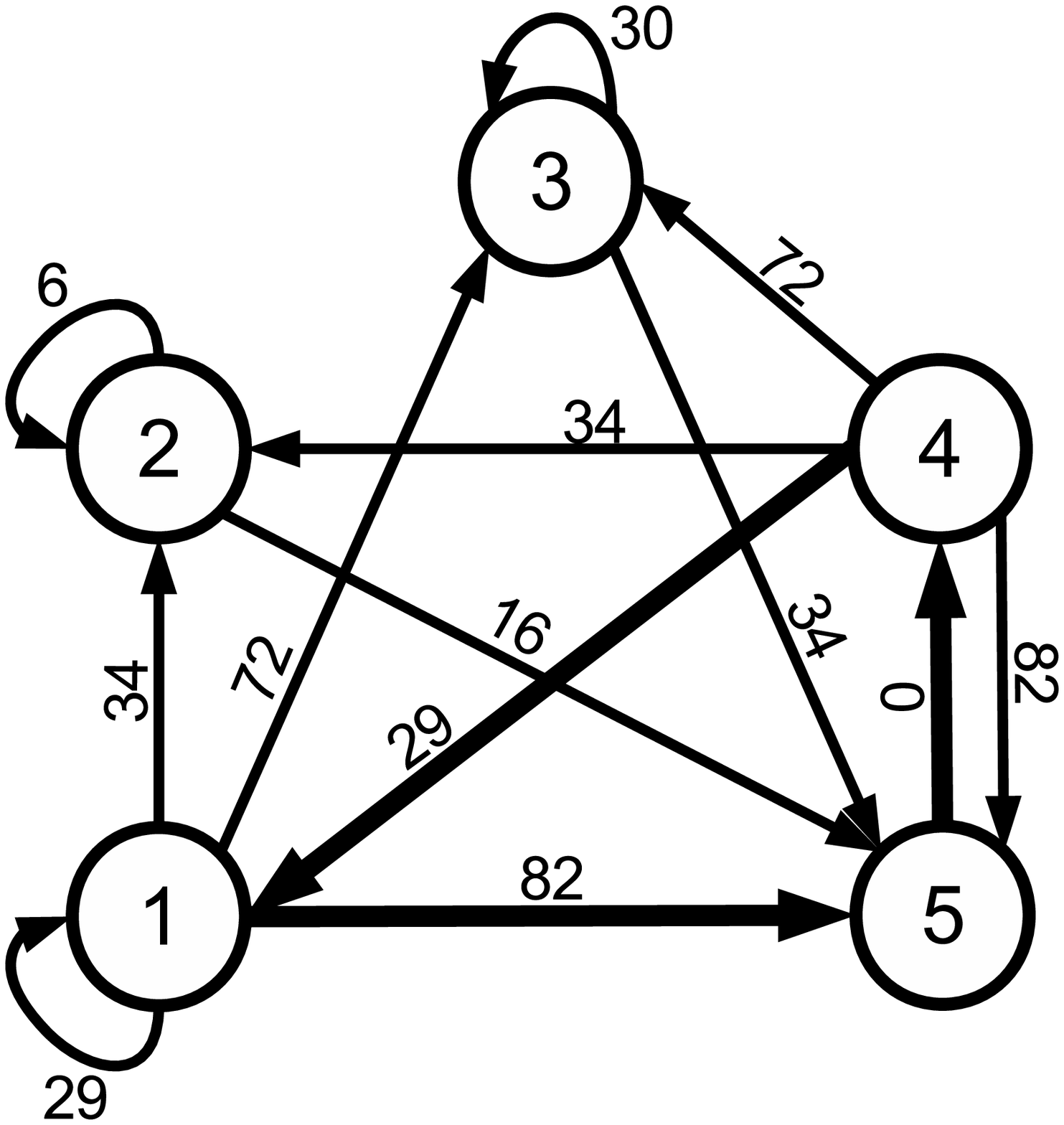}}
	\caption{Scenario-aware dataflow}%
	\label{fig:sadf}%
\end{figure}

SADF models the dynamism of an application in terms of modes or scenarios. Every scenario is modeled by an SDFG, while the occurence patterns of scenarios are given by an FSM. We give the following definition of an SADFG.
\begin{mydef}
A Scenario-aware dataflow graph (SADFG) is a tuple \\ $\mathit{SADFG}=(S, F)$, where:
\begin{itemize}
\item $S=\{s_i \; | \; s_i= (\mathit{scen}_i,G_i)\}$ is a set of ordered pairs of scenarios and their corresponding SDFGs;
%
%
%
\item $F=(Q,q_0, \delta, \Sigma, E)$ is the scenario finite state machine consisting of a finite set $Q$ of states, an initial state $q_0 \in Q$, a transition relation $\delta \subseteq Q \times Q$, a scenario labelling $\Sigma:  Q \rightarrow S$ and a set of final states $E$, where $E=Q$.
\end{itemize}
\end{mydef}
Fig. \ref{fig:example_ssdf} shows an example SADFG with two scenarios, $a$ and $b$. In this example both scenarios use the same scenario graph, but the actor execution times differ. For example, actor $A$ has a firing duration of $29$ time-units in scenario $a$ and $28$ time units in scenario $b$. The scenario FSM is fully connected and thus allowing arbitrary scenario order.

Every finite path of arbitrary length $\overline{q}$ over the FSM corresponds to a sequence $\overline{s}$ with $\overline{s}(k)=\Sigma(\overline{q}(k))$. When the FSM performs a transition, the SDFG graph associated with the destination state is executed for exactly one iteration. Let $\mathbf{G}(s_i)$ denote the $n \times n$ (max,+) characteristic matrix for the scenario $s_i$, where $n$ is the number of initial tokens in the SADFG. Then the completion time of a $k$-long sequence of scenarios can then be defined as a sequence of (max,+) matrix multiplications $\mathcal{A}(s_1 \ldots s_k)=\mathbf{G}(s_k)\ldots \mathbf{G}(s_1)\vec{i}$, where $\vec{i}$ specifies the initial enabling times of the graph's initial tokens and usually $\vec{i}=\vec{0}$. The worst case increase of $\mathcal{A}(\overline{s})$ for a growing length of $\overline{s}$ specifies the worst-case throughput for any sequence of scenarios \cite{2gaub} \cite{2geil:all}.
Reference \cite{2geil:all} explains how to build the MPAG of an SADFG. Again, the inverse of the MCM ($1/ \lambda$) of the obtained MPAG denotes the worst-case throughput of that particular SADFG. A special case that arises in practice, which will be of the utmost importance in our SADF parametrization, is when scenarios can occur in arbitrary order, yielding the SADF FSM to be fully connected and with a single state for each scenario. In that case, the throughput of an SADFG equals to the maximum cycle mean of the MPAG that corresponds to the (max,+) matrix $\mathbf{G} = \underset{q \in Q}{\mathit{max}} \left(\mathbf{G}(\Sigma(q)) \right)$ \cite{2geil:all}. The operator $\mathit{max}$ denotes taking the maximum of the elements of the individual scenario matrices. The corresponding scenario matrices for the example SADFG in Fig. \ref{fig:example_ssdf} are:
\[ \mathbf{G}(a) = \begin{bmatrix}
29 & -\infty & -\infty & 29 & -\infty \\
33 & 4 & -\infty & 33 & -\infty\\
63 & -\infty & 30 & 63 & -\infty\\
-\infty & -\infty & -\infty & -\infty & 0\\
64 & 5 & 31 & 64 & -\infty\\
\end{bmatrix}
\quad
\mathbf{G}(b) = \begin{bmatrix}
28 & -\infty & -\infty & 28 & -\infty \\
34 & 6 & -\infty & 34 & -\infty\\
72 & -\infty & 24 & 72 & -\infty\\
-\infty & -\infty & -\infty & -\infty & 0\\
82 & 16 & 34 & 82 & -\infty\\
\end{bmatrix}.\]
The critical cycle of the corresponding MPAG obtained from the maximized matrix $\mathbf{G}=\mathit{max}(\mathbf{G}(a),$ $ \mathbf{G}(b))$, is denoted by bold arrows in Fig. \ref{fig:example_sadf_mpag}. Throughput in this case equals $1/37$ iterations per time-unit. This example also demonstrates that the worst-case throughput value cannot simply be obtained by only considering the `worst-case' scenario, or by analysing the graph where each actor takes its worst-case execution time over all scenarios.
\section{Parametric Rate and Actor Execution Time SADF Analysis}
We start this section by formally defining the PSADF model and showing the (max,+) equivalence between SADF and PSADF. We use this result in defining the PSADF worst-case throughput calculation problem as a constrained optimization problem over the PSADF graph (PSADFG) parameter space, where the objective functions are elements of the symbolic PSADFG (max,+) characteristic matrix. We conclude by giving the theoretical foundation and the algorithm for symbolic PSADFG (max,+) characteristic matrix extraction. 
\subsection{Motivation and Model Definition}
SADF becomes impractical or even infeasible when it faces applications with a vast set of possible behaviours. We overcome this limitation by \textit{parametrization}. The problem of parametrization of a dataflow model in terms of rates is not an easy task as it raises questions about properties like liveness, boundedness and schedulability. A naive approach in just declaring any rate of interest as parametric, could render the graph to deadlock, be unbounded or unschedulable. Therefore we start from SPDF \cite{2frad:all}. The liveness and boundedness properties for SPDF are decidable. SPDF extends SDF by allowing rates to be parametric while preserving static schedulability. Rates are products of static natural numbers and/or parameters that can change dynamically. The changes of each parameter $p$ are made by a single actor called its modifier each $\alpha^\mathrm{th}$ time it fires using `$\mathit{set} \; p[\alpha]$' annotation. We re-define SPDF \cite{2frad:all} by adding the notion of time of SDF/SADF to it.
\begin{mydef}
A schedulable parametric dataflow graph (SPDFG) is a tuple  $\mathit{SPDFG}=(\mathcal{G}, \mathcal{PR}, \mathcal{PD}, i,$ $ r, e, M, \alpha)$, where:
\label{def:psadf}
\begin{itemize}
\item $\mathcal{G}$ is a directed connected graph $(\mathcal{A},\mathcal{E})$ with $\mathcal{A}$ set of actors and $\mathcal{E} \subseteq \mathcal{A} \times \mathcal{A}$ set of edges (\textit{channels});
\item $\mathcal{PR}$ is a set of rate parameters (symbolic variables) used to define SPDF rates by the grammar $ \mathcal{FR} ::= k \mid pr \mid \mathcal{FR}_1 \cdot \mathcal{FR}_2 $, where $pr \in \mathcal{PR}$, $ k \in \mathbb{N}^+$;
\item $\mathcal{PD}$ is a set of actor execution time parameters (symbolic variables) used to define SPDF actor execution times by the grammar $ \mathcal{FD} ::= k \cdot pd \mid \mathcal{FD}_1+\mathcal{FD}_2$, where $pd \in \mathcal{PD}$, $ k \in \mathbb{R}^+_0$;
\item $i:\mathcal{E} \rightarrow \mathbb{N}_{0}$ returns for each edge \textit{channel} its number of initial tokens;
\item $r:\mathcal{A} \times \mathcal{E} \rightarrow \mathcal{FR}$ returns for each port (represented by an actor and one of its edges) its rate;
\item $e:\mathcal{A} \rightarrow \mathcal{FD}$ returns for each actor its execution time;
\item $ M:\mathcal{PR} \rightarrow \mathcal{A}$ and $\alpha: \mathcal{PR} \rightarrow \mathcal{FR}$ returns for each rate parameter its modifier and its change period.
\end{itemize}
\end{mydef}
We consider only live SPDFGs as defined in \cite{2frad:all}. We allow parameters (rates and actor execution times) to change in between iterations. The introduction of parametric actor execution times to SPDF does not influence the liveness property. We define actor execution times as linear combinations of parameters. This gives us the ability to encode dependence, e.g. in case two actors are mapped onto the same processor, the ratio of their execution times will always be constant within an iteration.

Fig. \ref{fig:example_spdf} shows an example of a SPDF graph where actors have parametric ($p,q,s$) or constant rates and parametric execution times ($a,b,c,d,e$). Parametric rates $p$ and $s$ are modified by the actor $A$ every time it fires, while the parametric rate $q$ is modified by the actor $B$ every $p^\mathrm{th}$ time it fires.

Now we can define our parametric SADF model, by subjecting SPDF to the operational semantics of SADF.
\begin{mydef}
A parametric rate and parametric actor execution time SADFG (PSADFG) is a tuple  $\mathit{PSADFG}=(G, \Omega, F)$, where:
\begin{itemize}
\item $G$ is a live SPDFG;
\item $\Omega = \{ \vec{p} \; \mid \; \vec{p} \; \in \; {\mathbb{N}^{+}}^{|\mathcal{PR}|} \times  {\mathbb{R}^{+}_{0}}^{|\mathcal{PD}|} \}$ is a bounded and closed set of all allowed parameter values (rates and actor execution times) for $G$ or shortly the parameter space;
%
%
\item $F=(Q,{q_0}, \delta, \Sigma)$ is the scenario state transition system consisting of a possibly infinite set $Q$ of states, an initial state ${q_0} \in Q$, a transition relation $\delta \subseteq Q \times Q$ and a scenario labelling $\Sigma:  Q \rightarrow \Omega$.
\end{itemize}
\end{mydef}
In contrast to SADF, which explicitly defines scenarios as a finite collection of SDF graphs, in PSADF scenarios are implicitly defined over the bounded and closed vector parameter space $\Omega$. Elements of $\Omega$ are vectors $\vec{p} \; \in \; {\mathbb{N}^{+}}^{|\mathcal{PR}|} \times {\mathbb{R}^{+}_{0}}^{|\mathcal{PD}|}$. Let $\mathbf{G}(\vec{p})$ be the PSADF (max,+) characteristic $n\times n$ matrix for the parameter space point $\vec{p}$, where $n$ is the number of initial tokens in PSADFG.
The operational semantics of the model is as follows: every finite path of arbitrary length $\overline{q}$ over the scenario transition system $F$ corresponds to a sequence $\overline{s}$ with $\overline{s}(k)=\Sigma(\overline{q}(k))$. This is a sequence of parameters space points, i.e. $\overline{s}=\overline{\vec{p}}$. The evaluation of the PSADFG's SPDFG $G$ at a parameter space point is nothing else but an SDFG. The characteristic (max,+) matrix of this SDFG equals to $\mathbf{G}(\vec{p})$ (evaluation at a concrete $\vec{p} \in \Omega$). When the scenario state transition system performs a transition, the SDFG obtained by the evaluation of the PSADFG at that exact point is executed for exactly one iteration. Given previous reasoning, the analogy to SADF is obvious. We can say that PSADF is a compact representation of SADF. From the performance analysis perspective, by using the provision of an infinite (max,+) automaton \cite{2gaub} we can define the completion time of a $k$-long sequence of parameter point activations as a sequence of (max,+) matrix multiplications $\mathcal{A}(\overline{\vec{p}})=\mathbf{G}(\vec{p}_k)\ldots\mathbf{G}(\vec{p}_1)\vec{i}$ as it is done in \cite{2geil:all} for SADF. The worst case increase of $\mathcal{A}(\overline{\vec{p}})$ for a growing length of $\overline{\vec{p}}$ represents the worst-case throughput for any sequence of parameters points allowed by the scenario transition system.
%
%
%

As already  mentioned, PSADF is a compact representation of SADF. We use it to model the behaviour of applications characterized by vast number of scenarios where it is impossible to determine the scenario occurrence pattern even if such exists. Therefore, in terms of PSADF we will be considering the case of a fully connected scenario state transition system, i.e. $\delta = Q \times Q$, and where every state of the transition system corresponds to one parameter space point, i.e. there is a bijective mapping $z:Q \rightarrow \Omega$. This way we will always be able to give a conservative bound on the worst-case throughput. This is due to the simple fact that the language recognized by an arbitrary PSADF $F$ is always included in the language recognized by the PSADF $F$ where $\delta = Q \times Q$ and there exists a bijection $z:Q \rightarrow \Omega$.
\begin{myprop}
\label{prop:wct}
The worst-case throughput of a PSADFG for which $\delta = Q \times Q$ and for which exists a bijective mapping $z:Q \rightarrow \Omega$ equals to the inverse of the maximum cycle mean of the MPAG defined by the matrix $\mathbf{G}=\underset{q \in Q}{\mathit{max}}\left( \mathbf{G} (z(q))\right)$.
\begin{proof}
Given the operational semantics of PSADF previously described and the fact that $\Omega$ is bounded and closed, it follows straightforwardly from \cite{2geil:all}\cite{2gaub}.
\end{proof}
\end{myprop}
\subsection{Worst-Case Throughput Analysis}
\subsubsection{Problem Definition.}
Given $\mathbf{G}(\vec{p})=\{g_{ij}(\vec{p})\}$ as a matrix of continuous function over the closed and bounded parameter space $\Omega$ that possesses an appropriate mathematical formulation, e.g. as equalities and inequalities over a certain $\left( |\mathcal{PR}| + |\mathcal{PD}| \right)$-dimensional vector space, using Proposition \ref{prop:wct}, our worst-case throughput calculation problem becomes a set of maximally ($n\times n$) constrained optimization problems with $\mathbf{G}(\vec{p})=\{g_{ij}(\vec{p})\}$ as the objective function(s) and $\Omega$ as the constraint set:
\\\\
$\textbf{foreach} \;\; (i,j) \;\; \mbox{s.t.} \;\; g_{ij}(\vec{p})\ne -\infty \;\; \textbf{do}$
\begin{equation*}
\begin{aligned}
& \underset{\vec{p}}{\mbox{maximize}}
& & g_{ij}(\vec{p}) \\
&\mbox{subject to} 
& & \vec{p} \in \Omega.
\end{aligned}
\end{equation*}
A continuous function over a bounded and closed set admits a maximum. Of course, the term continuous includes also discrete functions that are continuous in the Heine sense. After maximizing all the element functions of $\mathbf{G}(\vec{p})$, the worst-case throughput will equal to the MCM of the MPAG given by the maximized PSADFG (max,+) characteristic matrix. Our main challenge is thus to derive a technique for the analytical formulation of the symbolic PSADFG (max,+) characteristic matrix $\mathbf{G}(\vec{p})$. $\mathbf{G}(\vec{p})$ is a matrix of functions that in the (max,+) sense encodes the time distances between initial tokens in adjacent iterations of a PSADFG. We will show that this is a matrix of polynomial functions of $\vec{p}$. Polynomial functions are continuous. Then the problem can be solved as a polynomial programming problem over $\Omega$. There exists a variety of techniques for solving such problems depending on the `shape' of $\Omega$. Do note here that these optimization problems are solved independently as we are interested in the worst-case increase of $\mathcal{A}(\overline{\vec{p}})$ for a growing length of $\overline{\vec{p}}$ (over a growing number of iterations).
\begin{figure}[t]%
\centering
	\subfloat[][An example SPDFG/PSADFG.]{
	\label{fig:example_spdf}
	\includegraphics[width=2.2in]{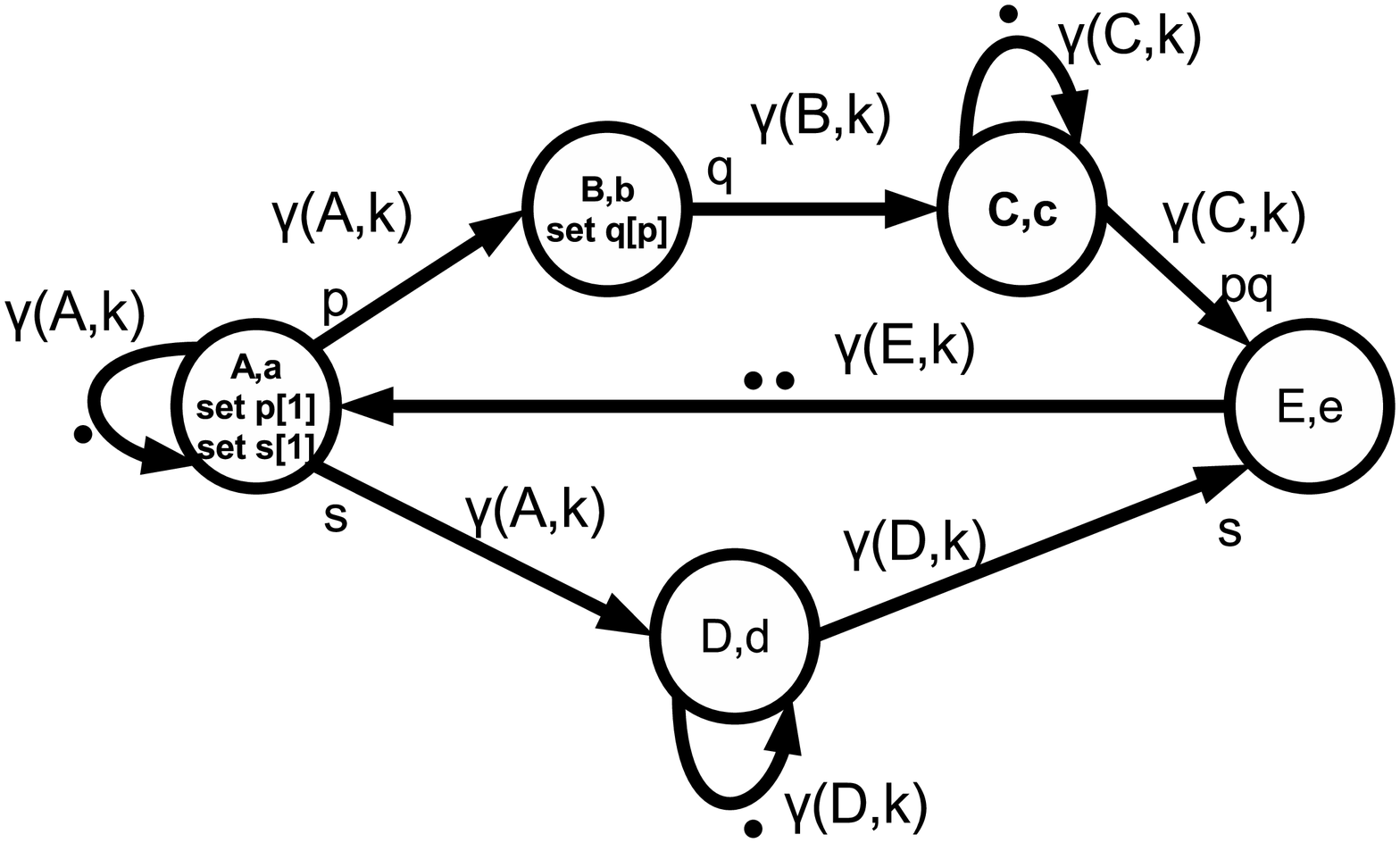}}%
	~~~~~~~~~~~~~~~
	\subfloat[][PSADF actor model.]{
	\label{fig:example_psadf_actor}
	\includegraphics[width=1.5in]{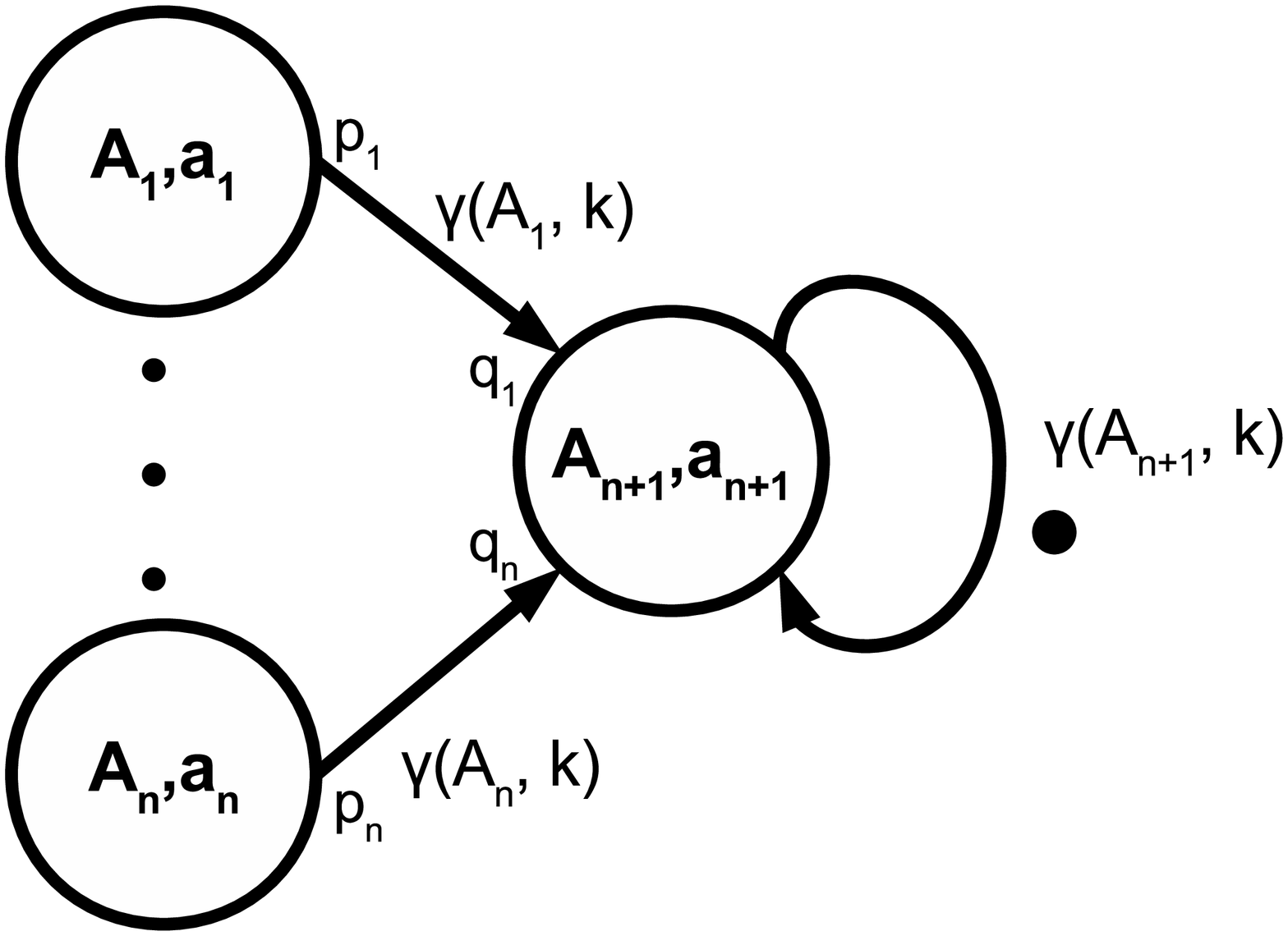}}
	\caption{Parametric SADF}%
	\label{fig:psadf}%
\end{figure}
\subsubsection{(max,+) Algebra for PSADF.}
In PSADF we only allow parameters to change between graph iterations, i.e $\frac{\#M(pr_j)}{\alpha(pr_j)}=1$ for parametric rates in the context of SPDF. The same goes for parametric actor execution times. Currently, our $\mathbf{G}(\vec{p})$ extraction technique requires that the considered PSADFG is `acyclic within an iteration'. If we take a PSADFG and convert it to a directed acyclic graph (PSADFG-DAG) by removing the edges with initial tokens, we require that only the PSADFG-DAG sink actors can produce tokens on the removed edges, and only the PSADFG-DAG source actors can consume from those edges. We do not include self-edges in this restriction. That is to say that we only allow cyclic dependencies tied to one actor. However, we can still consider PSADFGs that are serial compositions of subgraphs that are `acyclic within an iteration' if the subgraph performs only one iteration during an iteration of the composite PSADFG. Our $\mathbf{G}(\vec{p})$ extraction process will depend on the PSADFG quasi-static schedule which can be obtained using the procedure from \cite{2frad:all}. Basically, the PSADFG-DAG is sorted topologically. Result of the topological sorting is a string of actors. For PSADFG in Fig. \ref{fig:example_spdf} this string equals to $\mathit{ABCDE}$. Now we replace every actor $X$ with $X^{\#X}$, where $\#X$ is the PSADFG repetition vector entry for actor $X$. For PSADFG in Fig. \ref{fig:example_spdf} the final quasi-static schedule takes the form $AB^pC^{pq}D^sE$.

We continue by giving an appropriate (max,+) model of the PSADF actor as displayed in Fig. \ref{fig:example_psadf_actor}. First let us briefly explain the (max,+) semantics of a dataflow actor firing. If $T$ is the set of tokens needed by an actor to perform its firing and for every $\tau \; \in \;  T$, $t_\tau$ is the time that token becomes available, then the starting time of the actor firing is given by $\bigoplus \limits_{\tau \; \in \; T} t_\tau$. If $d$ is the execution time of that actor then the tokens produced by the actor firing become available at $\bigoplus \limits_{\tau \; \in \; T} t_\tau + d$.
Now, let $\gamma(A_i,k)$ be the completion time of the $k^{\mathrm{th}}$ firing of actor $A_i$. This annotation is present in Fig. \ref{fig:example_spdf} for each of the actors. 
In order for an actor to fire, it must have all its input dependencies satisfied. We can now derive the expression for $\gamma$:
\begin{equation}
\gamma(A_i, k) = \left( \bigoplus \limits_{A_h \mid (A_h, A_i) \in \mathcal{E} } \gamma\left( A_h, \ceil[\Big]{ \frac{r(A_i, (A_h,A_i))k-i(A_h,A_i)}{r(A_h, (A_h,A_i))} } \right)  \right) \otimes e(A_i).
\label{eq:main}
\end{equation} 
The completion time of the $k^{\mathrm{th}}$ firing of actor $A_i$ corresponds to the maximal completion times of appropriately indexed firings of actors that feed its input edges $A_h \mid (A_h, A_i) \in \mathcal{E}$ increased by its own execution time $e(A_i)$. The quotient $ \ceil[\Big]{ \frac{r(A_i, (A_h,A_i))k-i(A_h,A_i)}{r(A_h, (A_h,A_i))} }$ is used to index the appropriate firing of the actors that feed its input edges. The $i(A_h,A_i)$ member in the nominator of the fraction accounts for initial tokens. Initial tokens have the semantics of the initial delay and form the initial conditions used to solve (max,+) difference equations, analogue to the initial conditions in classical linear difference (recurrence) equations. We comply with the liveness criteria from \cite{2frad:all} which among others requires that all SPDFG cycles are live, i.e. within a cycle there is an edge with initial tokens to fire the actor the needed number of times to complete an iteration, either a global one or a local one. Liveness and the `acyclic within an iteration' restriction render \eqref{eq:main} solvable and we can always obtain a solution for \eqref{eq:main} in terms of initial conditions. The analytical solution of a system of such (max,+) linear difference equations evaluated at the iteration boundary for every actor of the graph will exactly give us the needed symbolic PSADFG characteristic (max,+) matrix. We follow the order of actors from the quasi-static schedule. This guarantees that we respect data/resource dependencies. Element $X^{\#X}$ tells us that we have to solve \eqref{eq:main} for actor $X$ at $k=\#X$. The obtained solution is propagated to the next iteration of the algorithm. We continue until we reach the end of the quasi-static schedule. At this point we will obtain solutions for all actors in terms of dependence of their completion times at the iteration boundary on initial conditions. From these solutions we can then easily construct the symbolic PSADFG characteristic (max,+) matrix. 

Let us consider the PSADFG example in Fig. \ref{fig:example_spdf}.
We write down (max,+) equations for each actor (we omit the sign $\otimes$, i.e. $a\otimes b$ will be denoted as $ab$):
\begin{equation}
\gamma(A,k) = \left( \gamma(A,k-1) \oplus \gamma(E,k-2) \right)a =  a\gamma(A,k-1) \oplus a\gamma(E,k-2),
\label{eq:A}
\end{equation}
\begin{equation}
\gamma(B,k)= b \gamma(A, \ceil{\frac{k}{p}}),
\label{eq:B}
\end{equation}
\begin{equation}
\gamma(C,k) = \left( \gamma(B,\ceil{\frac{k}{q}}) \oplus \gamma(C,k-1) \right)c =  c\gamma(B,\ceil{\frac{k}{q}}) \oplus c\gamma(C, k-1),
\label{eq:C}
\end{equation}
\begin{equation}
\gamma(D,k) = \left( \gamma(A,\ceil{\frac{k}{s}}) \oplus \gamma(D,k-1) \right)d =  d\gamma(A,\ceil{\frac{k}{s}}) \oplus d\gamma(D, k-1),
\label{eq:D}
\end{equation}
\begin{equation}
\gamma(E,k) = \left( \gamma(C,pqk) \oplus \gamma(D,sk) \right)e =  e\gamma(C,pqk) \oplus e\gamma(D,sk).
\label{eq:E}
\end{equation}
The initial conditions are:
\begin{align}
\gamma(A,0)=t_1,\;
\gamma(D,0)=t_2,\;
\gamma(C,0)=t_3,\;
\gamma(E,-1)=t_4,\;
\gamma(E,0)=t_5.
\end{align}
We can now evaluate and solve them at an iteration boundary given by the sequential schedule $AB^pC^{pq}D^sE$.
Firing actor $A$ using \eqref{eq:A} with $k=1$ we obtain:
\begin{equation}
\gamma(A,1)=a\gamma(A,0) \oplus a\gamma(E,-1)=at_1 \oplus at_4.
\label{eq:fA}
\end{equation}
Firing $B^p$ using \eqref{eq:B} with $k=p$ and using \eqref{eq:fA} we obtain:
\begin{equation}
\gamma(B,p)=abt_1 \oplus abt_4.
\label{eq:fB}
\end{equation}
Firing $C^{pq}$ using \eqref{eq:C} with $k=pq$ and \eqref{eq:fB} we obtain (backward substitution):
\begin{equation}
\gamma(C, pq)= abct_1 \oplus abct_4 \oplus c\gamma(C, pq-1)=abc^{pq}t_1 \oplus c^{pq}t_3 \oplus abc^{pq}t_4.
\label{eq:fCfinal}
\end{equation}
Firing $D^s$ using \eqref{eq:D} with $k=s$ similarly evaluates to:
\begin{equation}
\gamma(D, s)=ad^st_1 \oplus d^st_2  \oplus ad^st_4.
\label{eq:fD}
\end{equation}
Firing $E$ using \eqref{eq:E} with $k=1$ and \eqref{eq:fCfinal} \eqref{eq:fD} we obtain:
\begin{equation}
\gamma(E,1)= aet_1(bc^{pq} \oplus d^s) \oplus d^set_2 \oplus c^{pq}et_3 \oplus aet_4(bc^{pq} \oplus d^s).
\label{eq:fE}
\end{equation}
In \eqref{eq:fE} initial conditions $t_1$ and $t_4$ are (max,+) multiplied by a symbolic (max,+) summation term $(bc^{pq} \oplus d^s)$. We refer to this situation as a \textit{conflict}. The production time of the tokens generated by actor $E$ will depend on the relationship between $(b+pqc)$ and $sd$. Before proceeding, we have to consider two cases. One given by $(b+pqc \geq sd)$ and the other by $(b+pqc \le sd)$. We must check the intersection of newly added constraints and the already existing ones to reason against feasibility. If there are no feasible points in one of the subregions, we drop the further evaluation within the same subregion. In this example let us assume that both subregions contain feasible points. We easily construct the symbolic matrices from the solutions that are all expressed in terms of their dependence on initial conditions at an iteration boundary.
We write down once more the solutions of the equations at the iteration boundary for actors that reproduce the initial tokens. Those are actors $(A,C, D, E)$. We will change the notation from $\gamma(A_i,k)$ to $t_j^\prime$ depending on the indexes of initial conditions (tokens) and the producing actor. We obtain for $(b+pqc \geq sd)$:
\begin{align}
\label{eq:matrixf}
t_1^\prime&=at_1 \oplus at_4, \\
t_2^\prime&=ad^st_1 \oplus d^st_2  \oplus ad^st_4, \\
t_3^\prime&=abc^{pq}t_1 \oplus c^{pq}t_3 \oplus abc^{pq}t_4, \\
t_4^\prime&=t_5, \\
t_5^\prime&=abc^{pq}et_1 \oplus d^set_2 \oplus c^{pq}et_3 \oplus abc^{pq}et_4.
\label{eq:matrixl}
\end{align}
From \eqref{eq:matrixf}-\eqref{eq:matrixl} we then easily obtain the rows of the symbolic \textit{(max,+)} matrix:
\[ \mathbf{G}_{(b+pqc \geq sd)}= \begin{bmatrix}
a & -\infty & -\infty & a & -\infty \\
a+sd & sd & -\infty & a+sd & -\infty\\
a+b+pqc & -\infty & pqc & a+b+pqc & -\infty\\
-\infty & -\infty & -\infty & -\infty & 0\\
a+b+pqc+e & sd+e & pqc+e & a+b+pqc+e & -\infty\\
\end{bmatrix}.\]
The same procedure is used for the  $(b+pqc \le sd)$ case.
The evolution of the PSADF graph is then governed by the following equations over the parameter space $\Omega$: $\vec{\gamma}_{k+1}=\mathbf{G}_{(b+pqc \geq sd)} \vec{\gamma}_{k}$ and $\vec{\gamma}_{k+1}=\mathbf{G}_{(b+pqc \le sd)} \vec{\gamma}_{k}$, depending in which region of $\Omega$ is the $(k+1)^{\mathrm{th}}$ iteration scheduled. If $(b+pqc = sd)$, any of the two can be chosen. In the definition of both regions we use the $\le$ and $\ge$ operators to have them remain closed. The functions that constitute the symbolic (max,+) matrices are polynomial functions of $\vec{p}$.

In order to obtain the worst case throughput we will have to solve a mixed-integer polynomial programming problem for $\mathbf{G}_{(b+pqc \geq sd)}$ and $\mathbf{G}_{(b+pqc < sd)}$ over $(\Omega \cap (b+pqc \geq sd))$ and $(\Omega \cap (b+pqc \leq sd))$, respectively. A collection of techniques that solve such problems for a variety of definitions of $\Omega$, e.g. convex, non-convex or restricted to take only a few discrete values, can be found in \cite{2sher:all}. The matrix $\mathit{max}\left(\mathbf{G}_{(b+pqc \geq sd)}, \mathbf{G}_{(b+pqc \le sd)}\right)$ will define the MPAG of the example PSADFG. The inverse of the MCM of this MPAG equals to the worst-case throughput.

At this point we present our recursive algorithm for symbolic PSADF (max,+) characteristic matrix extraction (Algorithm \ref{alg:syme}).
\begin{algorithm}[!ht]
\caption{Symbolic PSADFG (max,+) characteristic matrix extraction}\label{alg:syme}
\begin{algorithmic}[1]
\Function{SymbolicExtract}{$\mathit{Qss,MpEqSet,\varPhi,Ss}$}
	\State $ \mathit{fBranchingNode \leftarrow fals}e$
	\While{$ \mathit{not \; Qss.isFinished()} $} \label{line:sched}
		\State $\mathit{currQssElem \leftarrow Qss.popNextElem()}$
		\State $\mathit{currSol} \leftarrow \Call{Solve}{$$\mathit{MpEqSet,currQssElem}$$}$ \label{line:solve}
		
		\If{$\mathit{currSol.Conflicted()}$}
			\State $\mathit{fBranchingNode \leftarrow true} $ \label{line:branch}
			\While{$\mathit{ new\varPhi \leftarrow currSol.getNextConflict()}$} 	\label{line:branch1}
				\If{\Call{FeasibilityCheck}{$\mathit{new\varPhi, \varPhi} $}} \label{line:feas}
					\State $\mathit{ curr\varPhi \leftarrow \varPhi}$
					\State $\mathit{curr\varPhi.Add(new\varPhi)} $	\label{line:addnewc}
					\State $\mathit{currMpEqSet \leftarrow MpEqSet}$
					\State $\mathit{currMpEqSet.ResolveC(new\varPhi)} $	\label{line:resolve}
					\State $\mathit{Ss}.Add(\Call{SymbolicExtract}{$$\mathit{Qss,currMpEqSet,curr\varPhi,Ss}$$)}$ \label{line:branchs}
				\EndIf
			\EndWhile
		\Else
			\State $\mathit{MpEqSet.Update(currSol)} $ \label{line:addsole}
		\EndIf
	\EndWhile\label{schedulewhile}
	\If{$\mathit{not \; fBranchingNode} $}
		\State  \Return $\mathit{(mpEqSet, \varPhi)}$ \label{line:end}
	\Else \State \Return $\emptyset$ \label{line:bla}
	\EndIf
\EndFunction
\end{algorithmic}
\end{algorithm}
\nopagebreak
The inputs to the algorithm are the pre-computed sequential quasi-static schedule $\mathit{Qss}$, the set of PSADF (max,+) difference equations $\mathit{MpEqSet}$, the initial parameter space $\varPhi=\Omega$ and the initial solution set $\mathit{Ss}=\emptyset$. The solution set $\mathit{Ss}$ is a set of ordered pairs $\mathit{Ss}=\{(\mathbf{G}_{\varPhi_i}, \varPhi_i)\}$, where $\mathbf{G}_{\varPhi_i}$ is the symbolic (max,+) matrix that governs the evolution of the PSADF in the region $\varPhi_i \subseteq \Omega$ generated by adding conflict resolving constraints to $\Omega$ during the execution of the algorithm.
Algorithm traverses the sequential schedule taking one actor with its repetition count at a time (Line \ref{line:sched}). Function \textsc{Solve} (Line \ref{line:solve}) solves Equation \eqref{eq:main} for the considered actor. If there are no conflicts in the solution, the algorithm updates the equation set with the current solution that can be used in later iterations (Line \ref{line:addsole}). If there are conflicts, i.e. there are $\bigoplus \limits_{i} y_i$ terms multiplying the initial conditions, we have to split the parameter space (Line \ref{line:branch1}). For example, if the term $y_1\oplus y_2 \oplus y_3$ is multiplying an initial condition, we have to consider three cases: $(y_1 > y_2, y_1 > y_3)$, $(y_2 > y_1, y_2 > y_3)$ and $(y_3 > y_1, y_3 > y_2)$. Function \textsc{FeasibilityCheck} (Line \ref{line:feas}) checks the emptiness of the intersection of the current constraint set $\varPhi$ and the new constraints. If the intersection is non-empty, new constraints are added to the current set for this branch of exploration (Line \ref{line:addnewc}), conflicts are resolved (Line \ref{line:resolve}) and \textsc{SymbolicExtract} is recursively called again (Line \ref{line:branchs}). If the intersection is non-feasible, this branch is dropped. If we continue in this fashion we will eventually reach a non-branching node (Line \ref{line:end}).

We demonstrate our approach on the example PSADF graph in Fig. \ref{fig:example_spdf}. The example models a dynamic streaming application consisting of loops with interdependent parametric affine loop bounds. We define the ranges for parametric loop bounds (PSADF rates) as: $p \in \left[10, 2000\right], q \in \left[10, 15\right]$ and $s \in \left[100, 1500\right]$. We also define linear dependencies between them: $p+s \le 1400$ and $q \le p$. Our application is run on a multi-processor platform where each loop body (actor) is mapped onto a different processor. Let PSADF actor execution times take the values of their nominal execution times multiplied by the parameter $c_i \in \left[1, 5\right]$ to account for six different possible platform dynamic voltage and frequency scaling (DVFS) settings. We obtain: $a=30c_i,\; b= 20c_i,\; c=4c_i,\; d=3c_i,\; e=c_i$. These constraints define $\Omega$ for our example. To obtain the worst-case throughput value we must maximize the matrices $\mathbf{G}_{(b+pqc \geq sd)}$ and $\mathbf{G}_{(b+pqc \le sd)}$ over $\Omega$ as given by the previously listed constraints. These become two mixed integer polynomial programming problems over $\Omega \cap (b+pqc \geq sd)$ and $\Omega \cap (b+pqc \le sd)$ and can be solved using the technique from \cite{2sher:all}. Throughput is given by the inverse of the MCM of the MPAG defined by the matrix $\mathit{max}\left(\ \mathbf{G}_{(b+pqc \le sd)}, \mathbf{G}_{(b+pqc \geq sd)}\right)$ and equals to $1/390000$ iterations per time-unit.
\section{Experimental results}
We demonstrate our throughput analysis technique on five representative DSP applications with parametric interdependent affine loop bounds listed in Table \ref{tabl:result}. The first column shows the number of PSADFG actors, the second denotes the number of initial tokens, the third shows the number of parametric rates, the fourth gives the number of parametric actor execution times and the last shows the number of scenarios as the number of points in the PSADFG parameter space $\Omega$. All applications, except the bounded block parallel lattice reduction algorithm for MIMO-OFDM \cite{2ahmad:all}, are mapped onto a two-processor scalar architecture. The latter is mapped onto a vector/SIMD architecture. To obtain the nominal actor execution times for our benchmark set, we used the AVR32 \cite{2avr} simulator under a reference frequency of 32 $\mathrm{MHz}$. For bounded block parallel lattice reduction algorithm \cite{2ahmad:all} we used random numbers for nominal actor execution times, as the source code of the algorithm is not publicly available. We assume that the frequency of each platform processor can be placed inside the range from 32 to 64 $\mathrm{MHz}$, with the step of 1 $\mathrm{Mhz}$. For a 2 processor platform this will give 32 possible combinations. In contrast to the conventional SADF approach from \cite{2geil:all} which would have to generate $|\Omega|$ SDFGs, our approach in each of these cases will solve maximally ($n \times n$) polynomial programming problems without the need for the enumeration  of $\Omega$ which is a difficulty by itself. Actually, in practice this number is usually less than ($n \times n$), because not all initial tokens depend on all other initial tokens in the graph rendering the matrices to be quite sparse. Moreover, sometimes the entries in the symbolic PSADF (max,+) characteristic matrix are repetitive, so we only have to solve the corresponding problem once. The symbolic PSADF (max,+) characteristic matrices of the benchmark applications were extracted manually using Algorithm \ref{alg:syme}, while the corresponding optimization problems were solved using CVX, a package for specifying and solving convex programs \cite{cvx}\cite{gb08}.
\begin{table}
\caption{Experimental results}
\begin{center}
\label{tabl:result}
\begin{tabular}{c|c|c|c|c|c}
\hline
\noalign{\smallskip}
Benchmark & $ |\mathcal{A}|$ & $n$ & $|\mathcal{PR}|$ & $|\mathcal{PD}|$ & $|\Omega|$ \\
\noalign{\smallskip}
\hline
Fundam. freq. detector based on norm. autocorr. \cite{2icts:all} & $12$ & $6$ & $2$ & $2$ & $16,687,681 \cdot 32$ \\
Normalized LMS alg. \cite{2icts:all} & $9$ & $6$ & $2$ & $2$ & $385 \cdot 32$\\
High resolution spectral analysis \cite{2icts:all} & $9$ & $6$ & $2$ & $2$ & $385 \cdot 32$\\
Adaptive predictor program \cite{2chass} & $6$ & $4$ & $2$ & $2$ & $400 \cdot 32$\\
Bound. block parallel latt. reduct. alg. \cite{2ahmad:all} & $12$ & $5$ & $3$ & $1$ & $300 \cdot 16$ \\
\hline
\end{tabular}
\end{center}
\end{table}
\section{Conclusion}
In this paper we have presented an extension to SADF that allows to model applications with vast or infinite sets of behaviours. We refer to our model as PSADF. We have proven the semantical equivalence of the two models and used that result in the formulation of worst-case throughput calculation problem for PSADF graphs with a fully connected state transition system within a generic optimization framework. The objective functions are functionals that represent the elements of the symbolic PSADF (max,+) characteristic matrices. Furthermore, we have derived a (max,+) linear theory based algorithm that is able to generate these matrices by combining a (max,+) difference equation solver and a recursive parameter space exploration for a subclass of PSADF graphs that are `acyclic within an iteration'. As future work, we want to fully automate our technique and investigate the problem of parametric throughput analysis of PSADF graphs.

\nocite{*}
\bibliographystyle{eptcs}
\bibliography{syncop}
\end{document}